\documentclass[final,5p,times,twocolumn,number]{elsarticle}

\biboptions{sort&compress}

\usepackage{amssymb}
\usepackage{lipsum}
\usepackage{amsmath}
\usepackage{xcolor}
\usepackage{amsthm}

\journal{Physics Letters B}

\begin{document}

\begin{frontmatter}



\title{Flavour Invariants of the N Higgs Doublet Model}


\author[first]{João C. Belas}
\author[first]{João P. Silva}
\affiliation[first]{organization={CFTP, Departamento de Física, Instituto Superior Técnico, Universidade de Lisboa},
            addressline={Avenida Rovisco Pais 1},
            postcode={1049},
            city={Lisboa},
            country={Portugal}}

\begin{abstract}
In this work, a systematic way of analysing the N Higgs Doublet Model
flavour sector will be developed.
We introduce a complete set of mixing matrices describing the rotation between
certain suitably defined bases, akin to the Cabibbo–Kobayashi–Maskawa matrix,
which describes the relation between the up-quark and down-quark mass bases.
We point out the crucial importance played by the charged Higgs basis.
A complete set of weak basis transformation invariant traces of flavour matrices
for the general N doublets case is also introduced for the first time.
This will be important for studies of the renormalization group evolution in terms
of relevant physical parameters.
\end{abstract}




\end{frontmatter}

%\tableofcontents



\section{\label{sec: intro} Introduction}
The Standard Model (SM) of Particle Physics has been well established by measurements
of unprecedented precision, including the 2012 discovery
of the first fundamental scalar: the Higgs boson \cite{ATLAS:2012yve,CMS:2012qbp}.
Nevertheless,
one knows that it cannot be the complete description of Nature:
it does not describe the 85\% dark matter content of the Universe;
it does not have the phase transition and CP violation strength necessary to account for the 
observed baryon asymmetry;
and, in its simplest form, does not account for neutrino masses.
Two other problems remain unsolved. On the one hand, having found one fundamental scalar,
it is of paramount importance to determine how many such scalars there exist in Nature.
This has led to a continuous theoretical and experimental effort studying
the effect of extra scalars, including the N Higgs Doublet Model (NHDM).
On the other hand, the hierarchy of fermionic masses and mixing angles
(the so-called flavour problem) lacks a satisfactory explanation.
In particular, models with extra doublets introduce a further twist to the flavour problem.
Indeed, in NHDM there are, in general, flavour changing neutral couplings
(FCNC) of scalars with fermions.
So far,
no systematic identification of the invariants
relevant for flavour physics in the NHDM has been developed.
This will be undertaken here.

The search of invariants under basis transformations has a long history dating back,
at least, to the development of the so-called Jarlskog invariant,
parametrizing the sole source of CP violation in the SM
\cite{Jarlskog:1985ht,Dunietz:1985uy,Botella:1985gb}.
Shortly thereafter,
more invariants probing CP violation in the gauge-fermion sector
were established for more than three families \cite{Bernabeu:1986fc,Gronau:1986xb},
for models with vectorlike quarks \cite{Branco:1986my},
for the Left-Right symmetric model \cite{Branco:1986pr},
for neutrinos with Majorana masses \cite{Branco:1986gr},
and for supersymmetric models \cite{Branco:1986kf}.

Extensions to models with more than one Higgs doublet were first proposed in
Ref.~\cite{Mendez:1991gp,Lavoura:1994fv,Botella:1994cs}.
In particular,
Ref.~\cite{Botella:1994cs} proposed a systematic method to find
CP-even and CP-odd basis invariant quantities.
In short, one must
build products of the coupling matrices appearing in the Lagrangian,
of increasing complexity, taking traces over all the (scalar or fermion) internal flavour
spaces. Since traces have been taken, these expressions are invariant under the
basis transformations. Those traces with an imaginary part signal CP violation.
This method has been extended to provide invariant quantities which signal
the breaking of R-parity in supersymmetric theories \cite{Davidson:1996cc}.
Interest in basis invariant quantities for the
two Higgs doublet model resurfaced after 2004
\cite{Ginzburg:2004vp,Davidson:2005cw,Gunion:2005ja,Ivanov:2005hg,
Haber:2006ue,Maniatis:2006fs,Ivanov:2007de,Nishi:2007dv}.
CP violation invariants for the scalar sector of the NHDM ($N \geq 3$)
were studied in \cite{Branco:2005em,Nishi:2006tg}.

These methods require care in identifying the independent
number of invariants needed; one starts by identifying how many parameters are required,
after basis transformations are taken into account, and then carefully determines what
specific invariants to take.
A very interesting specific example can be found in the 1988 Ref.~\cite{Branco:1987mj},
where the ten invariants describing the quark masses and mixings of the SM are
clearly identified.
It was also the procedure used in Ref.~\cite{Botella:2012ab} in order to study flavour
in the 2HDM.
Other recent articles using this or similar techniques,
mostly in the context of CP,
include \cite{Lebedev:2002wq,Botella:2004ks,Dreiner:2007yz,
deMedeirosVarzielas:2016rii,Yu:2019ihs,Yu:2020gre,
Darvishi:2023ckq,Darvishi:2024cwe}.
This is the method applied in this article.

More recently,
a mathematical technique involving 
Hilbert series and the plethystic logarithm has been developed,
in order to identify a theory's invariants.
This method permits the counting and identification of primary invariants,
and also of relations with further invariants (syzygies).
This technique has been extensively applied to a variety of
models, for example, in
\cite{Benvenuti:2006qr,Feng:2007ur,Jenkins:2009dy,Hanany:2010vu,
Lehman:2015via,Henning:2015daa,Lehman:2015coa,Henning:2015alf,
Henning:2017fpj,Bednyakov:2018cmx,Trautner:2018ipq,
Bento:2020jei,Bento:2021hyo,Yu:2021cco,Bonnefoy:2021tbt,
Yu:2022ttm,Bento:2023owf,Grinstein:2023njq,
deLima:2024vrn}.
Using this procedure,
there are two types of invariants (primary and secondary);
a classification which does not exist in the previous method.
This is not the method followed in this article.

In section \ref{sec: SM Yukawa Lagrangian}, the flavour sector of the SM is analysed. 
The Yukawa Lagrangian of models with extra scalar doublets is presented in 
section \ref{sec: Multi Higgs Yukawa Lagrangian}, as well as the notions 
of the \textit{Higgs basis} \cite{Georgi:1978ri,Donoghue:1978cj,Botella:1994cs}
and the \textit{charged Higgs basis} \cite{Nishi:2007nh,Bento:2017eti,Bento:2018fmy}.
In section \ref{sec: The idea}, a new way of looking at the flavour sector of 
the NHDM is introduced, in terms of new mixing matrices and interaction eigenvalues. 
Section \ref{sec: Physical Parameters} analyses the NHDM in light of 
this new perspective, taking into account the physically relevant parameters.
This section is fully devoted to the intricate issue of parameter counting
and matching to the number of invariants.
Section \ref{sec: Weak Basis Transformations} explains the notion of 
\textit{Weak Basis Transformations} (WBT) and how to build invariant
quantities under such operations.
These are then used in section \ref{sec: Flavour Invariants}
to build a complete set of flavour invariants for the NHDM in light
of this new way of analysing flavour physics.
The conclusions are presented in section \ref{sec:Conclusions}.
\ref{sec: unitary matrix} contains some useful definitions
regarding the parametrisation of unitary matrices.
And finally 
\ref{app: NHDM} simplifies some expressions obtained in
section \ref{sec: Flavour Invariants}.

\section{\label{sec: SM Yukawa Lagrangian} Standard Model Yukawa Lagrangian}
In the SM, there is only one scalar doublet and the quarks' Yukawa Lagrangian is
\begin{equation}
-\mathcal{L}_Y =  \overline{Q_L}\Gamma\Phi n_R
+ \overline{Q_L}\Delta \widetilde{\Phi} p_R 
+ h.c.  \;,
\end{equation}
where $\overline{Q_L} =(\overline{p_L}, \overline{n_L})$ are the $SU(2)$ quark left-handed 
doublets,
$p_R$ ($n_R$) are the charge +2/3 (-1/3) quark right-handed singlets,
$\Gamma$ and $\Delta$ are generic 3 by 3 complex Yukawa matrices,
$\widetilde{\Phi} = i\sigma_2\Phi^*$, $\sigma_2$ is the second Pauli matrix,
and $h.c.$ stands for the hermitian conjugate.
After spontaneous symmetry breaking (SSB),
the scalar field acquires a vacuum expectation value (vev) $v/\sqrt{2}$.
The quarks mass matrices are
\begin{equation}
M_d = \frac{v}{\sqrt{2}}\Gamma\, ,
\; \; \; \; \; \; \; \;
M_u = \frac{v}{\sqrt{2}}\Delta\, .
\end{equation}
One can rotate the left-handed and right-handed quark fields
\begin{eqnarray}
n_L = V_{dL}d_L, &\ \ & n_R = V_{dR}d_R\, ,
\nonumber\\
p_L = V_{uL}u_L, &\ \ & p_R = V_{uR}u_R\, ,
\end{eqnarray}
into a basis
where their interaction with the Higgs field is diagonal,
\begin{eqnarray}
M_d &=& V_{dL} D_d V_{dR}^\dag, \; \; \; \; D_d = \textrm{diag}\{m_d, m_s, m_b\}\, ,
\nonumber\\
M_u &=& V_{uL} D_u V_{uR}^\dag, \; \; \; \; D_u = \textrm{diag}\{m_u, m_c, m_t\}\, ,
\end{eqnarray}
thus performing the singular value decomposition (SVD) of $M_u$ and $M_d$.
In this basis, the Yukawa Lagrangian relevant to the quark masses becomes
\begin{eqnarray}
-\mathcal{L}_Y &\supset& \overline{d_L}V_{dL}^\dag M_d V_{dR}d_R
+ \overline{u_L}V_{uL}^\dag M_u V_{uR}u_R  + h.c.
\nonumber\\
&=&
\overline{d_L}D_d d_R + \overline{u_L}D_u u_R  + h.c. \; .
\end{eqnarray}
Since the left chiral up-type and down-type quarks are part of doublets,
there are 2 relevant bases for these doublets:
the basis where the up-type quarks are mass eigenstates,
and the basis where the down-type quarks are mass eigenstates.
The failure of these two bases to coincide is encoded in
the Cabibbo–Kobayashi–Maskawa (CKM)
matrix \cite{Cabibbo:1963yz,Kobayashi:1973fv}, as
\begin{equation}
    V = V_{uL}^\dag V_{dL}.
\end{equation}
In order to count the number of parameters, it is useful to look at the
SVD of the mass matrices $M_d$ and $M_u$.
Taking $M_d$ as an example, it is a general 3 by 3 complex matrix,
thus having 9 magnitudes and 9 phases.
In contrast, the diagonal matrix $D_d$ has 3 magnitudes and each
unitary matrix $V_{dL}$ and $V_{dR}$ has 3 magnitudes (angles) and 6 phases,
which can be parametrised as in
Eqs.~\eqref{unitary parametrisation} and \eqref{U2} of the appendix. 
Thus, the magnitudes match, but there are 3 redundant
phases. Its SVD can be written explicitly as
\begin{eqnarray}
M_d &=& \{1,\alpha_1^L,\alpha_2^L\}\textbf{R}_{23}^L\{-\delta^L,1,1\}
\textbf{R}_{13}^L\{\delta^L,1,1\}\textbf{R}_{12}^L
\nonumber\\
&&
\{\alpha_3^L,\alpha_4^L,\alpha_5^L\} \textrm{diag}\{m_d,m_s,m_b\}
\{-\alpha_3^R,-\alpha_4^R,-\alpha_5^R\}
\nonumber\\
&&
\textbf{R}_{-12}^R\{-\delta^R,1,1\}\textbf{R}_{-13}^R\{\delta^R,1,1\}
\textbf{R}_{-23}^R\{1,-\alpha_1^R,-\alpha_2^R\},
\nonumber \\
&&
\hspace{1cm}
\end{eqnarray}
where the $L(R)$ superscript comes from the $V_{dL}(V_{dR})$ parameters.
Computing the matrix product explicitly,
one can see that the phases $\alpha_3^L$ and $\alpha_3^R$ only
appear in $M_d$ as the linear combination $\alpha_3^L-\alpha_3^R$,
and likewise for $\alpha_4$ and $\alpha_5$.
As such, for each of these $\alpha$'s,
one can pick one of them to be determined by the entries
of $M_d$ and the other one to be a free arbitrary parameter.
It will be useful to choose $\alpha_i^L$ to be the arbitrary ones.
The same line of thinking can be applied to the up sector.
With such a choice of arbitrary $\alpha$'s,
the CKM matrix becomes explicitly
\begin{eqnarray}
V &=& \{-\alpha_3^u,-\alpha_4^u,-\alpha_5^u\}
\nonumber\\
&&
\textbf{R}_{-12}^u\{-\delta^u,1,1\}\textbf{R}_{-13}^u\{\delta^u,1,1\}\textbf{R}_{-23}^u\{1,-\alpha_1^u,-\alpha_2^u\}
\nonumber\\
&&
\{1,\alpha_1^d,\alpha_2^d\}\textbf{R}_{23}^d\{-\delta^d,1,1\}\textbf{R}_{13}^d\{\delta^d,1,1\}\textbf{R}_{12}^d
\nonumber\\
&&
\{\alpha_3^d,\alpha_4^d,\alpha_5^d\}.
\label{V_general}
\end{eqnarray}
Here, the notation was simplified, by dropping the redundant $L$ superscript and replacing it 
with either $u$ or $d$, for the up-type and down-type sector parameters, respectively.
The two middle rows of Eq.~\eqref{V_general} constitute
themselves a unitary matrix, and, as such,
can be reparametrised following Eq.~\eqref{U2}, leading
to~\footnote{Here the angles $\theta_{ij}$ ($1 \leq i<j \leq3$)
in $\textbf{R}_{ij}$ and the phase $\delta$ are the specific measured angles 
and phase of the CKM matrix.
This is in contrast to the expressions in the appendices, where the angles
and phases are generic, referring to the parametrisation of a generic matrix.}
\begin{eqnarray}
V &=& \{-\alpha_3^u,-\alpha_4^u,-\alpha_5^u\}
\nonumber\\
&&
\{1,\alpha_1,\alpha_2\}\textbf{R}_{23}\{-\delta,1,1\}
\textbf{R}_{13}\{\delta,1,1\}\textbf{R}_{12}\{\alpha_3,\alpha_4,\alpha_5\}
\nonumber\\
&&
\{\alpha_3^d,\alpha_4^d,\alpha_5^d\}.
\end{eqnarray}
Since the $\alpha^d$'s and $\alpha^u$'s were left as arbitrary,
one can choose them to cancel the other 5 $\alpha$'s,
leaving, at last, the familiar parametrisation of the CKM matrix,
with 3 mixing angles
(from $\textbf{R}_{23}$, $\textbf{R}_{13}$, and $\textbf{R}_{12}$)
and 1 complex phase ($\delta$). 

Therefore, in the SM, the minimum number of parameters required to
describe the quark Yukawa part of its Lagrangian is 10:
6 quark masses (3 up-type and 3 down-type),
3 CKM mixing angles,
and 1 CKM CP violating phase.

\section{\label{sec: Multi Higgs Yukawa Lagrangian}N Higgs Doublet Model's Yukawa Lagrangian}
In the general case containing N scalar doublets,
after SSB,
one can parametrise them as
\begin{equation}
    \Phi_k
    = e^{i\alpha_k}\left( \begin{array}{c} \phi^+_k \\
\frac{1}{\sqrt{2}}(v_k + \rho_k + i \eta_k)
\end{array} \right),
\; \; \; \; k = \{1, ..., N\}.
\end{equation}
The quarks' Yukawa Lagrangian generalises to
\begin{equation}
-\mathcal{L}_Y =  \overline{Q_L}\Gamma_k\Phi_k n_R
+ \overline{Q_L}\Delta_k \widetilde{\Phi}_k p_R 
+ h.c.  \, .
\end{equation}
There is an implicit summation over $k$,
which will continue to be used throughout this section.
The $\Gamma_k$ and $\Delta_k$ are general 3 by 3 complex matrices
that mediate the interaction between the $k^{th}$ scalar doublet and
the negatively and positively charged quarks, respectively.

The relevant matrices for the quarks' masses become
\begin{equation}
M_d = \frac{1}{\sqrt{2}} \left(e^{i\alpha_k}v_k\Gamma_k\right),
\; \; \; \; \; \; \; \;
M_u = \frac{1}{\sqrt{2}} \left(e^{-i\alpha_k}v_k\Delta_k\right).
\end{equation}
And, in order to diagonalise them,
the same SVD process used for the SM can be applied here.

\subsection{The Higgs Basis and the charged Higgs basis}

It is useful to perform a basis rotation of the N scalar doublets
into the so-called \textit{Higgs basis},
where the vev is isolated into a single doublet
\begin{eqnarray}
\left( \begin{array}{c}  \mathcal{H}_0
\\ ... \\
\mathcal{H}_{N-1}  \end{array} \right)
 =
\mathcal{S} \left( \begin{array}{c} e^{-i\alpha_1}\Phi_1
\\ ... \\
e^{-i\alpha_N}\Phi_N \end{array} \right),
\end{eqnarray}
and $\mathcal{S}$ is a unitary matrix such that the Higgs doublets become
\begin{eqnarray}
\mathcal{H}_0
&=& \left( \begin{array}{c} G^+ \\
\frac{1}{\sqrt{2}}(v + H^0 + i G^0) \end{array} \right),
\nonumber\\
\mathcal{H}_k
&=& \left( \begin{array}{c} H^+_k \\
\frac{1}{\sqrt{2}}(R_k + i I_k) \end{array} \right),
\; \; \; \; k = \{1, ..., N - 1 \},
\end{eqnarray}
where $v = \sqrt{v_1^2 + ... + v_N^2}$ and $\mathcal{S}_{1j} = \frac{v_j}{v}$.
With this choice, the $G^\pm$ and $G^0$ would-be Goldstone bosons will, in the Unitary gauge,
be absorbed as the longitudinal degree of freedom of the
$W^\pm$ and $Z^0$ bosons, respectively.
The boson $H^0$ has the same Yukawa and gauge couplings
as the SM Higgs boson,
despite possibly not being a mass eigenstate.
At this stage, there is great redundancy in how to define the remaining rows
of $\mathcal{S}$, the corresponding doublets and, thus, the
corresponding Yukawa matrices.

To reach physically meaningful parameters,
it is convenient to choose the matrix $\mathcal{S}$ such that the fields
$H^+_k $ are already the charged scalar mass eigenstates; 
this is known as the \textit{charged Higgs basis}.
After the transformation to the charged Higgs basis,
the full Yukawa Lagrangian becomes \cite{Bento:2018fmy}
\begin{eqnarray}
-\mathcal{L}_Y
&=&
\left( \overline{d_L}V_{dL}^\dag M_d V_{dR} d_R
+ \overline{u_L} V_{uL}^\dag M_u V_{uR} u_R\ \right)
\nonumber\\
&&
+ \frac{H^0}{v} \left(  \overline{d_L} V_{dL}^\dag M_d V_{dR} d_R
+ \overline{u_L} V_{uL}^\dag M_u V_{uR} u_R \right)
\nonumber\\
&&
+ \frac{R_k}{v} \left( \overline{d_L} V_{dL}^\dag N_{dk} V_{dR} d_R
+  \overline{u_L} V_{uL}^\dag N_{uk} V_{uR} u_R  \right)
\nonumber\\
&&
+ \frac{iI_k}{v} \left( \overline{d_L} V_{dL}^\dag N_{dk} V_{dR} d_R
- \overline{u_L} V_{uL}^\dag N_{uk} V_{uR} u_R \right)
\nonumber \\
&&
+ \frac{\sqrt{2}H^+_k }{v}
\left( \overline{u_L} V_{uL}^\dag N_{dk} V_{dR} d_R
- \overline{u_R} V_{uR}^\dag N_{uk}^\dag V_{dL} d_L \right)
\nonumber \\
&&
+ h.c.\, 
\nonumber \\
&=&
\left( \overline{d_L}D_d d_R +  \overline{u_L} D_u u_R\ \right)
\nonumber\\
&&
+ \frac{H^0}{v} \left(  \overline{d_L} D_d d_R + \overline{u_L} D_u u_R \right)
\nonumber\\
&&
+ \frac{R_k}{v} \left( \overline{d_L} V_{dL}^\dag N_{dk} V_{dR} d_R
+  \overline{u_L} V_{uL}^\dag N_{uk} V_{uR} u_R \right)
\nonumber\\
&&
+ \frac{iI_k}{v} \left( \overline{d_L} V_{dL}^\dag N_{dk} V_{dR} d_R
- \overline{u_L} V_{uL}^\dag N_{uk} V_{uR} u_R \right)
\nonumber \\
&&
+ \frac{\sqrt{2}H^+_k }{v}
\left( \overline{u_L} V_{uL}^\dag N_{dk} V_{dR} d_R
- \overline{u_R} V_{uR}^\dag N_{uk}^\dag V_{dL} d_L \right)
\nonumber \\
&&
+ h.c.\, , 
\end{eqnarray}
where $N_{dk}$, and $N_{uk}$ depend on the specific form of $\mathcal{S}$, as
\begin{eqnarray}
\left( \begin{array}{c} M_d \\ N_{d1}
\\ ... \\
N_{dN-1}  \end{array} \right)
&=&
v \; \mathcal{S} \left( \begin{array}{c} e^{i\alpha_1}\Gamma_1 \\
e^{i\alpha_2} \Gamma_2
\\ ... \\
e^{i\alpha_N} \Gamma_N \end{array} \right), 
\nonumber\\
\left( \begin{array}{c} M_u \\ N_{u1}
\\ ... \\
N_{uN-1}  \end{array} \right)
&=&
v \; \mathcal{S} \left( \begin{array}{c} e^{-i\alpha_1} \Delta_1 \\
e^{-i\alpha_2} \Delta_2
\\ ... \\
e^{-i\alpha_N} \Delta_N \end{array} \right).
\end{eqnarray}
As in the SM case,
the rotation matrices
$V_{dL}$, $V_{dR}$, $V_{uL}$, and $V_{uR}$
will diagonalise the $M_d$ and $M_u$ matrices,
giving the Lagrangian in the mass basis.
The FCNC are mediated by the matrices
$N_{dk}$ and $N_{uk}$, which are, in general, non diagonal.
It is also possible to further transform the neutral scalar bosons
into their mass eigenstates,
but for the purposes of this work that will not be necessary. 

Regarding the scalar sector, before SSB, there is a global $U(N)$ basis freedom, 
since the scalar doublets are all in the same representations of the SM gauge group and can 
rotate among themselves. After SSB and the rotation to the Higgs basis, there is a $U(N-1)$ 
global basis freedom, since in that case one of the doublets is special, $\mathcal{H}_0$, which 
contains the vev (again chosen to lie on the real direction of the neutral component 
of the doublet), but all the other doublets can be rotated among themselves. 
After a rotation to the charged Higgs basis, assuming that the charged Higgs 
boson's masses are not degenerate, the  global $U(N-1)$ is broken down to a global 
$U(1)^{N-1}$, since now each $\mathcal{H}_k$ doublet 
contains a physical charged Higgs mass eigenstate. However, each doublet,
but the one containing the vev,
can still be rephased, and the resulting basis will still be a charged Higgs basis. 
Therefore, the charged Higgs basis is defined up to a rephasing of each of the $\mathcal{H}_k$ 
doublets.

\section{\label{sec: The idea}A New Way to Look at Flavour Physics
in the NHDM}
As there is a special basis (the mass basis) in which 
the quarks'
interaction with $\mathcal{H}_0$ is diagonal,
achieved through rotation matrices $V_{dL}$, $V_{dR}$, $V_{uL}$, $V_{uR}$,
there is also a basis in which the quarks' interaction
with $\mathcal{H}_k, k \in \{1, ..., N - 1 \}$ is diagonal,
achieved through rotation matrices $U_{dLk}$, $U_{dRk}$, $U_{uLk}$, $U_{uRk}$.
Indeed, just as the SVD of $M_d$ and $M_u$ was performed to arrive at
$D_d$ and $D_u$, one can perform the SVD of $N_{dk}$ and $N_{uk}$,
as~\footnote{Notice that, in general, $(d_1,d_2,d_3)$ are not the
mass eigenstates $(d,s,b)$, and similarly for the up sector.}
\begin{eqnarray}
N_{dk} &=& U_{dLk} E_{dk} U_{dRk}^\dag, \; \; \; \; E_{dk}
= \textrm{diag}(n_{d_1 k}, n_{d_2 k}, n_{d_3 k}),
\nonumber\\
N_{uk} &=& U_{uLk} E_{uk} U_{uRk}^\dag, \; \; \; \; E_{uk}
= \textrm{diag}(n_{u_1 k}, n_{u_2 k}, n_{u_3 k}).
\nonumber\\
&&
\label{SVD_N}
\end{eqnarray}
The $E_{dk}$ and $E_{uk}$ entries correspond to the eigenvalues
of the interaction between the three generations of quarks of each
sector and $\mathcal{H}_k$ in this new basis, in particular with the charged Higgs' mass
eigenstates.
The special case of the 2HDM was discussed in \cite{Botella:2012ab},
that arrived at the SVD decomposition of the $N_d$ and $N_u$ matrices.
But, in that particular case, there is only one FCNC matrix for each quark sector.
In contrast, when there are three (or more) scalar doublets, there are two (or more)
$N$ matrices in each sector, with great redundancy in which two matrices to take.
What gives physical meaning to our choice is the notion of charged Higgs basis;
in that particular basis, all $N$ matrices acquire physical significance.
Of course, in the 2HDM the Higgs basis is already a charged Higgs basis
(since there is only one massive charged Higgs),
and we recover the result in \cite{Botella:2012ab}.

There are three types of quark fields: $Q_L$, $n_R$, and $p_R$.
In the case of the SM, there is one relevant basis for each $n_R$, $p_R$,
their respective mass basis, and two relevant bases for $Q_L$,
the basis in which $n_L$ are mass diagonal and the basis in which $p_L$ are mass diagonal.
The CKM matrix measures the difference between these two latter bases.
In the NHDM case, there are $N$ relevant bases for each $n_R$, $p_R$,
the mass basis and the $N - 1$ new ``\textit{interaction with $\mathcal{H}_k$}'' bases,
and $2N$ relevant bases for ${Q_L}$, the ${n_L}$ mass basis, the ${p_L}$ mass basis,
the $N - 1$ ${n_L}$ ``\textit{interaction with $\mathcal{H}_k$}'' bases,
and the $N - 1$ ${p_L}$ ``\textit{interaction with $\mathcal{H}_k$}'' bases.
As such, there will be $4N - 3$ physical change of bases matrices:
$N - 1$ relating the $N$ relevant bases for $n_R$;
$N - 1$ relating the $N$ relevant bases for $p_R$;
and $2N - 1$ change of bases matrices for the $2N$ relevant bases of $Q_L$.
A natural way to construct these is to compute the difference in bases
between the mass basis and the ``\textit{interaction with $\mathcal{H}_k$}'' basis,
for each of the $n_L$, $p_L$, $n_R$, and $p_R$ fields, with the last rotation matrix
being the familiar SM CKM matrix relating the up-type and down-type left chiral quarks
\begin{eqnarray}
&&
C_{dLk} = V_{dL}^\dag U_{dLk},
\; \; \; \; C_{uLk} = V_{uL}^\dag U_{uLk},
\nonumber\\
&&
C_{dRk} = V_{dR}^\dag U_{dRk}, \; \; \; \;
C_{uRk} = V_{uR}^\dag U_{uRk},
\nonumber\\
&&
V = V_{uL}^\dag V_{dL},
\end{eqnarray}
since this gives a special emphasis to the mass basis. 
In \cite{Botella:2012ab}, these matrices were used in the 2HDM case
to compute CP violating flavour invariants, as will be explained in
more detail in section \ref{sec: Flavour Invariants}.
For the NHDM, there will be $N - 1$ matrices of each type, adding to $4N - 4$ which,
when including the CKM matrix, gives the correct number of $4N - 3$.
These $C$ matrices encode the FCNC of the model,
since they account for the failure of the quarks mass basis
and ``\textit{interaction with $\mathcal{H}_k$}'' basis to be the same.
If all these matrices were to be diagonal, or the eigenvalues of all
the $E_k$ matrices were to vanish, there would be no FCNC.

In this new way of looking at flavour physics,
the Yukawa Lagrangian becomes, in the mass basis,
\begin{eqnarray}
-\mathcal{L}_Y
&=&
\left( \overline{d_L}D_d d_R +  \overline{u_L} D_u u_R\ \right)
\nonumber\\
&&
+ \frac{H^0}{v} \left(  \overline{d_L} D_d d_R
+ \overline{u_L} D_u u_R \right)
\nonumber\\
&&
+ \frac{R_k}{v} \left( \overline{d_L} C_{dLk} E_{dk} C_{dRk}^\dag d_R
+  \overline{u_L} C_{uLk} E_{uk} C_{uRk}^\dag u_R \right)
\nonumber\\
&&
+ \frac{iI_k}{v} \left( \overline{d_L} C_{dLk} E_{dk} C_{dRk}^\dag d_R
-  \overline{u_L} C_{uLk} E_{uk} C_{uRk}^\dag u_R \right)
\nonumber \\
&&
+ \frac{\sqrt{2}H^+_k }{v}
\left( \overline{u_L} V C_{dLk} E_{dk} C_{dRk}^\dag d_R
\right.
\nonumber\\
&&
\hspace{17mm}
\left.
- \overline{u_R} C_{uRk} E_{uk} C_{uLk}^\dag V d_L \right)
%\nonumber \\
%&&
+\ h.c.\, .
\end{eqnarray}
Written in this way,
all parameters have physical significance (up to rephasings of the $\mathcal{H}_k$ doublets).
This result is decisive for the following sections.

This provides a systematic analysis of the flavour
sector of the NHDM, explaining it in terms of a number of relevant bases
for the quarks and rotation matrices between those bases.

\section{\label{sec: Physical Parameters}Physical Parameters}
The procedure applied to compute the number of physical parameters in the SM can
be modified to suit this new way of looking at flavour physics.
As occurs in the SM, the SVD of mass matrices $M_d$ and $M_u$ produces
diagonal matrices $D_d$ and $D_u$, left rotation matrices $V_{dL}$ and $V_{uL}$,
and right rotation matrices $V_{dR}$ and $V_{uR}$.
One can choose the 3 arbitrary phases in each sector to be in
$V_{dL}$ and $V_{uL}$, respectively,
and then use them to cancel the extra phases of the CKM matrix,
arriving at 10 physical parameters:
6 quark masses, 3 CKM mixing angles and 1 CKM CP violating phase.
Proceeding in this way,
these extra arbitrary phases become fixed,
and cannot be further used to remove more parameters.

The exact same procedure can be applied for models with more than one scalar doublet.
In addition, concerning the SVD of the $N_{dk}$ and $N_{uk}$ matrices,
one can choose the 3 arbitrary phases to be in either the left,
or right rotation matrices of each sector.
Therefore, one can choose them to be on the $U_{dLk}$ and $U_{uLk}$ matrices.
As such, the matrix $C_{dLk}$ becomes
\begin{eqnarray}
C_{dLk} &=& \{-\alpha_3^m,-\alpha_4^m,-\alpha_5^m\}
\nonumber\\
&&
\textbf{R}_{-12}^m\{-\delta^m,1,1\}\textbf{R}_{-13}^m
\{\delta^m,1,1\}\textbf{R}_{-23}^m\{1,-\alpha_1^m,-\alpha_2^m\}
\nonumber\\
&&
\{1,\alpha_1^k,\alpha_2^k\}\textbf{R}_{23}^k
\{-\delta^k,1,1\}\textbf{R}_{13}^k\{\delta^k,1,1\}\textbf{R}_{12}^k
\nonumber\\
&&
\{\alpha_3^k,\alpha_4^k,\alpha_5^k\},
\end{eqnarray}
where in the phase superscripts the $m$'s refer to the fact that the $V_{dL}$ matrix comes from
the SVD of $M_d$ and the $k$'s refer to the fact that the
$U_{dLk}$ matrix comes from the SVD of $N_{dk}$.
Since $\alpha_3^m$, $\alpha_4^m$, and $\alpha_5^m$ were fixed
in order for the CKM matrix to only have one complex phase,
the only arbitrary parameters in $C_{dLk}$ are $\alpha_3^k$, $\alpha_4^k$, and $\alpha_5^k$.
Then, since the first three rows are themselves a unitary matrix, they can be reparametrised as
\begin{eqnarray}
C_{dLk} &=&  \{1,\alpha_1^{dLk},\alpha_2^{dLk}\}\textbf{R}_{23}^{dLk}\{-\delta^{dLk},1,1\}
\nonumber\\
&&
\textbf{R}_{13}^{dLk}\{\delta^{dLk},1,1\}\textbf{R}_{12}^{dLk}
\{\alpha_3^{dLk},\alpha_4^{dLk},\alpha_5^{dLk}\}
\nonumber\\
&&
\{\alpha_3^k,\alpha_4^k,\alpha_5^k\}.
\end{eqnarray}
And, since the $\alpha^k$'s were left as arbitrary, one can
choose them to cancel $\alpha_3^{dLk}$, $\alpha_4^{dLk}$,
and $\alpha_5^{dLk}$, leaving the $C_{dLk}$ matrices with
3 mixing angles and 3 complex phases.
The same applies for $C_{uLk}$.
Since all arbitrary phases were used to reduce the number of
parameters in $V$, $C_{dLk}$, $C_{uLk}$, the matrices $C_{dRk}$ and $C_{uRk}$
have the maximal number of 3 mixing angles and 6 phases.

Regarding the scalar sector,
the effects of a  $U(1)^{N-1}$ transformation from one charged Higgs
basis to another by angles $\theta_k$ would be
\begin{equation}
    \mathcal{H}_k \rightarrow e^{i\theta_k}\mathcal{H}_k, \; \; \; \; N_{uk} \rightarrow e^{-i\theta_k}N_{uk}, \; \; \; \; N_{dk} \rightarrow e^{i\theta_k}N_{dk}.
\end{equation}
In terms of the parameters of the Yukawa Lagrangian, this would be equivalent to
\begin{equation}
    \alpha_j^{uRk} \rightarrow \alpha_j^{uRk} - \theta_k, \; \; \; \; \alpha_j^{dRk} \rightarrow \alpha_j^{dRk} + \theta_k, \; \; \; \; j = \{3, 4, 5\}.
\end{equation}
As such, for each $k$, one would think that out of these 6 parameters, only 5 combinations of 
them are physical.
However, these rephasings of the scalar doublets also rephase some parameters 
of the scalar potential, and one could also conclude that $k$ of the phases present in the 
scalar potential would be unphysical. In reality, out of all the parameters modified by the 
$U(1)^{N-1}$ transformations, both in the Yukawa interaction and scalar potential, only one for 
each $k$ becomes redundant.
Therefore,
one can choose to make $k$ parameters of the scalar potential redundant,
giving physical meaning to all of the Yukawa parameters described above.

Let us check to see whether this reasoning is consistent with the number
of physical parameters. Without the Yukawa sector, the Lagrangian would have a global $U(3)^3$
symmetry (a $U(3)$ for each $Q_L$, $n_R$, and $p_R$). After introducing the Yukawa sector, the
Lagrangian has a global $U(1)$, where all the fields are rephased by the same angle.
Following \cite{Santamaria_1993}, the physical number of parameters in
the Yukawa sector is given by
\begin{equation}
\label{eq: counting parameters}
    N_{Y\textrm{phys}} = N_{Y} - (N_G - N_{G'}),
\end{equation}
where $N_G$ and $N_{G'}$ are the initial and final global symmetry group generators. This 
equation works separately for magnitudes and phases. Naively, the Yukawa sector would introduce
$18N$ magnitudes and $18N$ phases (9 of each for each Yukawa matrix). Each global $U(3)$ 
symmetry has 3 magnitudes and 6 phases, and a $U(1)$ symmetry has 1 phase generator.
Therefore,
using Eq.~\eqref{eq: counting parameters},
there are $18N - 9$ physical magnitudes and 
$18N - (18 - 1) = 18N - 17$ physical phases.
Nine of the magnitudes and one phase correspond to the familiar SM Yukawa parameters,
which make a complete set in the SM case of $N = 1$.
The remaining $18(N - 1)$ both magnitudes and phases correspond to the
interaction with $\mathcal{H}_k$, given by $N_{dk}$ and $N_{uk}$ before the
rotation to the mass basis. Each of these two matrices is a general
3 by 3 complex matrix, giving the total of $18(N - 1)$
magnitudes and $18(N - 1)$ phases.

In this new way of looking at flavour physics, besides the 10 SM parameters,
there are:
$18(N - 1)$ extra magnitudes, $3(N - 1)$ from the diagonal
entries of $E_{dk}$ and $E_{uk}$ each, $3(N - 1)$ mixing angles from
each of the rotation matrices $U_{dLk}$, $U_{uLk}$, $U_{dRk}$, $U_{uRk}$;
and $18(N - 1)$ extra phases, $3(N - 1)$ complex phases from the
rotation matrices $U_{dLk}$, and $U_{uLk}$ each, and $6(N - 1)$ complex
phases from the rotation matrices $U_{dRk}$, and $U_{uRk}$ each.
As such, after considering the scalar $U(1)^{N-1}$ transformations and using the resulting
redundancy to eliminate $k$ parameters in the scalar potential, the number of parameters in the 
Yukawa sector is consistent.

\section{\label{sec: Weak Basis Transformations}Weak Basis Transformations}
A WBT is a transformation of the
fermion fields which leaves invariant
the gauge-kinetic terms \cite{Branco:1999fs}.
A certain quantity will be a flavour invariant if it is left unchanged
by a WBT,
in which the left handed quark doublets, the positive right handed
quark singlets, and the negative right handed quark singlets transform
independently by arbitrary unitary matrices $W_L$, $W_p$, and $W_n$, respectively.
Yukawa matrices will change under a WBT into
\begin{equation}
    \Delta_i' = W_L^\dag \Delta_iW_p, \; \; \; \; \Gamma_i' = W_L^\dag \Gamma_iW_n.
\end{equation}
Since the $M$ and $N$ matrices are linear combinations of the Yukawa
$\Delta$ and $\Gamma$ matrices, the same change happens for those matrices,
and it will be relevant to work in the charged Higgs basis henceforth.
In order to construct flavour invariants, one can multiply a Yukawa matrix
in the charged Higgs basis by the hermitian conjugate of another
(possibly different) Yukawa matrix of the same
sector~\footnote{There is an implicit dependence on $i,j$ in
$U_1$, $D_1$, etc., which for simplicity we do not make explicit.}
,
\begin{equation}
    U_1 = {N_u}_i{N_u^\dag}_j, \; \; \; \; D_1 = {N_d}_i{N_d^\dag}_j.
\end{equation}
These will transform as
\begin{eqnarray}
&&
U_1' = W_L^\dag{N_u}_iW_pW_p^\dag{N_u^\dag}_jW_L = W_L^\dag{N_u}_i{N_u^\dag}_jW_L,
\nonumber\\
&&
D_1' = W_L^\dag{N_d}_iW_nW_n^\dag{N_d^\dag}_jW_L = W_L^\dag{N_d}_i{N_d^\dag}_jW_L.
\end{eqnarray}
These matrices live on the left-handed doublet space,
both on the left and on the right side of the matrices,
as illustrated by the fact that they transform exclusively under $W_L$.
Indeed, by multiplying by hermitian conjugates, the spaces of right-handed
fields have effectively been traced over.
Likewise, one could construct matrices where both sides live
on the right-handed up-type (down-type) quark spaces, respectively as
\begin{equation}
    U_2 = {N_u^\dag}_i{N_u}_j, \; \; \; \; D_2 = {N_d^\dag}_i{N_d}_j,
\end{equation}
which will transform as
\begin{eqnarray}
&&
U_2' = W_p^\dag{N_u^\dag}_iW_LW_L^\dag{N_u}_jW_p = W_p^\dag{N_u^\dag}_i{N_u}_jW_p,
\nonumber\\
&&
D_2' = W_n^\dag{N_d^\dag}_iW_LW_L^\dag{N_d}_jW_n = W_n^\dag{N_d^\dag}_i{N_d}_jW_n.
\end{eqnarray}
Therefore, using the cyclic property of the trace, one can construct
flavour invariants from the trace of any of the preceding ``same-space''
matrices,
since (schematically)
\begin{eqnarray}
Tr(I') &=& Tr(W^\dag N_iN_j^\dag W) = Tr(WW^\dag N_iN_j^\dag)
\nonumber\\
&=& Tr(N_iN_j^\dag) = Tr(I).
\end{eqnarray}
Thus, in order to construct flavour invariants that connect the up and
down quark sectors, which will be relevant to recover the CKM parameters,
and in the NHDM with $N > 1$, some other phases too,
one can use the product of two matrices living on the left-space,
one constructed using Yukawa matrices from the up sector, and one
constructed using Yukawa matrices from the down sector.
The resulting matrix will still live on the left-space,
and its trace will be flavour invariant.

Finally, the rephasings of the $\mathcal{H}_k$ fields under the global $U(1)^{N-1}$ 
transformation must be taken into account. Since these rephasings also introduce an overall 
phase to the $N_{uk}$ and $N_{dk}$ matrices, in order for the results to be invariant under 
this $U(1)^{N-1}$ global symmetry,
\begin{equation}
    \#N_{uk} - \# N_{uk}^\dag - \# N_{dk} + \# N_{dk}^\dag = 0, 
\end{equation}
where $\#$ represents the number of times the respective matrix appears in a flavour invariant.

\section{\label{sec: Flavour Invariants}Flavour Invariants}
In the SM, matrices with both sides living on the left-space
can be constructed from the $M_u$, and $M_d$ mass matrices as
\begin{eqnarray}
&&
H_u = M_uM_u^\dag = V_{uL} D_u^2 V_{uL}^\dag,
\nonumber\\
&& 
H_d = M_dM_d^\dag = V_{dL} D_d^2 V_{dL}^\dag.
\end{eqnarray}
The 10 physical Yukawa parameters can be obtained from the flavour invariants
\cite{Branco:1987mj,Bento:2023owf},
\begin{eqnarray}
&&
I_1^0 = Tr(H_u), \; \; \; \; I_2^0 = Tr(H_u^2), \; \; \; \; I_3^0 = Tr(H_u^3),
\nonumber\\
&&
I_4^0 = Tr(H_d), \; \; \; \; I_5^0 = Tr(H_d^2), \; \; \; \; I_6^0 = Tr(H_d^3),
\nonumber\\
&&
I_7^0 = Tr(H_uH_d), \; \; \; \;  I_8^0 = Tr(H_u^2H_d), 
\nonumber\\
&&
I_9^0 = Tr(H_uH_d^2), \; \; \; \;  I_{10}^0 = Tr(H_u^2H_d^2).
\label{SM_10_invariants}
\end{eqnarray}
As can be seen from equations
\ref{simplification invariants SM} and \ref{invariants expression eigenvalues} of the appendix,
the invariants containing traces of $H_u$ ($H_d$) alone can be simplified to 
expressions composed exclusively of the quark masses. Since those algebraic expressions 
are all independent from each other and the invariants are equal in number to the parameters, 
3 up-type (down-type) quark masses, the invariants are independent from each other, and these 
parameters can be recovered from the values of the invariants.

Using equations \ref{simplification invariants SM} and \ref{invariants expression angles}, the 
invariants containing both $H_u$ and $H_d$ can be written in terms of the 
quark masses and CKM parameters. Since the quark masses have already been 
determined from the previous invariants, and since the algebraic expressions of the remaining 
invariants are all independent from each other and equal in number to the CKM parameters, these 
invariants are independent from each other and from the previous ones, and
these parameters can be computed from the remaining SM invariants.

A further invariant, 
whose magnitude is completely determined by the 10 invariants
in Eq.~\eqref{SM_10_invariants},
but whose sign carries new information,
was introduced by Jarlskog in order to probe directly CP violation
in a basis invariant way \cite{Jarlskog:1985ht}.

For an extended Yukawa Lagrangian with $N$ scalar doublets, it will be relevant to
define additional matrices
\begin{eqnarray}
&&
J_{uk} = N_{uk}N_{uk}^\dag = U_{uLk} E_{uk}^2 U_{uLk}^\dag,
\nonumber\\
&&
J_{dk} = N_{dk}N_{dk}^\dag = U_{dLk} E_{dk}^2 U_{dLk}^\dag, 
\nonumber\\
&&
A_{u} = M_{u}^\dag M_{u} = V_{uR} D_{u}^2 V_{uR}^\dag,
\nonumber\\
&&
A_{d} = M_{d}^\dag M_{d} = V_{dR} D_{d}^2 V_{dR}^\dag,
\nonumber\\
&&
B_{uk} = N_{uk}^\dag N_{uk} = U_{uRk} E_{uk}^2 U_{uRk}^\dag,
\nonumber\\
&&
B_{dk} = N_{dk}^\dag N_{dk} = U_{dRk} E_{dk}^2 U_{dRk}^\dag,
\nonumber\\
&&
F_{uk} = M_{u} N_{uk}^\dag = V_{uL} D_{u} C_{uRk} E_{uk} C_{uLk}^\dag V_{uL}^\dag,
\nonumber\\
&&
F_{dk} = M_{d} N_{dk}^\dag = V_{dL} D_{d} C_{dRk} E_{dk} C_{dLk}^\dag V_{dL}^\dag,
\nonumber\\
&&
G_{uk} = F_{uk}^\dag = N_{uk} M_{u}^\dag = U_{uLk} E_{uk} C_{uRk}^\dag D_{u} C_{uLk} U_{uLk}^\dag,
\nonumber\\
&&
G_{dk} = F_{dk}^\dag = N_{dk} M_{d}^\dag = U_{dLk} E_{dk} C_{dRk}^\dag D_{d} C_{dLk} U_{dLk}^\dag.
\end{eqnarray}
In \cite{Botella:2012ab}, several flavour invariants in the 2HDM were constructed,
mostly in the context of a systematic analysis of CP violation,
by building Jarlskog-like invariants using some of the
matrices defined above. 
Here, that reasoning is expanded,
by providing an extended list of invariants.
These invariants were chosen to allow the correspondence with the flavour
parameters of the NHDM. 
They will be simplified in terms of the underlying 
parameters in the appendix.

Using equations \ref{simplification invariants eigenvalues} and 
\ref{invariants expression eigenvalues} of the appendix, the same logic that was used to match 
some SM invariants to the quark masses can be used to match the following invariants to the 
eigenvalues of the interaction with $\mathcal{H}_k$
\begin{eqnarray}
&&
I_{1}^k = Tr(J_{uk}), \; \; \; \; I_{2}^k = Tr(J_{uk}^2), \; \; \; \; I_{3}^k = Tr(J_{uk}^3),
\nonumber\\
&&
I_{4}^k = Tr(J_{dk}), \; \; \; \; I_{5}^k = Tr(J_{dk}^2), \; \; \; \; I_{6}^k = Tr(J_{dk}^3),
\end{eqnarray}
for the up and down sectors, respectively, making these invariants independent from 
each other and from the previous ones. 

Using equations 
\ref{simplification invariants angles} and \ref{invariants expression angles} of the appendix, 
the same logic that was used to match the remaining SM invariants to the CKM parameters can be 
used to match the following invariants to the 3 mixing angles
and CKM-like phase of the matrices $C_{uLk}$, $C_{dLk}$, $C_{uRk}$,
and $C_{dRk}$
\begin{eqnarray}
&&
I_{7}^k = Tr(H_uJ_{uk}), \; \; \; \; I_{8}^k = Tr(H_u^2J_{uk}),
\nonumber\\
&&
I_{9}^k = Tr(H_uJ_{uk}^2), \; \; \; \; I_{10}^k = Tr(H_u^2J_{uk}^2),
\nonumber\\
&&
I_{11}^k = Tr(H_dJ_{dk}), \; \; \; \; I_{12}^k = Tr(H_d^2J_{dk}),
\nonumber\\
&&
I_{13}^k = Tr(H_dJ_{dk}^2), \; \; \; \; I_{14}^k = Tr(H_d^2J_{dk}^2),
\nonumber\\
&&
I_{15}^k = Tr(A_uB_{uk}), \; \; \; \; I_{16}^k = Tr(A_u^2B_{uk}),
\nonumber\\
&&
I_{17}^k = Tr(A_uB_{uk}^2), \; \; \; \; I_{18}^k = Tr(A_u^2B_{uk}^2),
\nonumber\\
&&
I_{19}^k = Tr(A_dB_{dk}), \; \; \; \; I_{20}^k = Tr(A_d^2B_{dk}),
\nonumber\\
&&
I_{21}^k = Tr(A_dB_{dk}^2), \; \; \; \; I_{22}^k = Tr(A_d^2B_{dk}^2),
\end{eqnarray}
respectively, making these invariants independent from each other and from the 
previous ones. 

As such, it has been established that the invariants introduced so far are all 
independent from each other and can be used in a closed form to obtain all the masses, 
eigenvalues, mixing angles, and CKM-like phase of the matrices introduced in section \ref{sec: 
The idea}. Therefore, one needs new independent invariants that also include the extra phases of 
the $C_{uLk}$, $C_{dLk}$, $C_{uRk}$, and $C_{dRk}$ matrices, in order to have a complete basis. 

Using equations 
\ref{simplification invariants left phases}, and \ref{invariants expression left phases 1}, and 
\ref{invariants expression left phases 2} of the appendix, one can write the following invariants 
in terms of the parameters already matched to previous invariants and the 2 extra complex phases 
of $C_{uLk}$, and $C_{dLk}$, respectively
\begin{eqnarray}
&&
I_{23}^k = Tr(J_{uk}H_d), \; \; \; \; I_{24}^k = Tr(J_{uk}^2H_d),
\nonumber\\
&&
I_{25}^k =  Tr(J_{dk}H_u), \; \; \; \; I_{26}^k = Tr(J_{dk}^2H_u).
\end{eqnarray}
Since the other parameters were already matched to the 
previous invariants, and since the algebraic expressions of these new invariants are all 
independent from each other and equal in number to the new parameters, these new invariants can 
be matched to the 2 extra complex phases of $C_{uLk}$, and $C_{dLk}$, respectively, making them 
independent from each other and from the previous ones. 

Using equations 
\ref{simplification invariants right phases}, and \ref{invariants expression right phases} of the 
appendix, one can write the following invariants in terms of the parameters already matched to 
previous invariants and the 5 extra complex phases of $C_{uRk}$, 
and $C_{dRk}$
\begin{eqnarray}
I_{27}^k
&=&
\frac{1}{2}\left(Tr(F_{uk}F_{dk}) + Tr(G_{dk}G_{uk})\right),
\nonumber\\
I_{28}^k
&=&
\frac{1}{2i}\left(Tr(F_{uk}F_{dk}) - Tr(G_{dk}G_{uk})\right),
\nonumber\\
I_{29}^k
&=&
\frac{1}{2}\left(Tr(H_uF_{uk}F_{dk}) + Tr(G_{dk}G_{uk}H_u)\right),
\nonumber
\end{eqnarray}
\begin{eqnarray}
I_{30}^k
&=&
\frac{1}{2i}\left(Tr(H_uF_{uk}F_{dk}) - Tr(G_{dk}G_{uk}H_u)\right),
\nonumber\\
I_{31}^k
&=&
\frac{1}{2}\left(Tr(H_dF_{dk}F_{uk}) + Tr(G_{uk}G_{dk}H_d)\right),
\nonumber\\
I_{32}^k
&=&
\frac{1}{2i}\left(Tr(H_dF_{dk}F_{uk}) - Tr(G_{uk}G_{dk}H_d)\right),
\nonumber\\
I_{33}^k
&=&
\frac{1}{2}\left(Tr(F_{uk}J_{uk}F_{dk}) + Tr(G_{dk}J_{uk}G_{uk})\right),
\nonumber\\
I_{34}^k
&=&
\frac{1}{2i}\left(Tr(F_{uk}J_{uk}F_{dk}) - Tr(G_{dk}J_{uk}G_{uk})\right),
\nonumber\\
I_{35}^k
&=&
\frac{1}{2}\left(Tr(F_{dk}J_{dk}F_{uk}) + Tr(G_{uk}J_{dk}G_{dk})\right),
\nonumber\\
I_{36}^k
&=&
\frac{1}{2i}\left(Tr(F_{dk}J_{dk}F_{uk}) - Tr(G_{uk}J_{dk}G_{dk})\right).
\end{eqnarray}
Since the other parameters were already matched to the 
previous invariants, and since the algebraic expressions of these new invariants are all 
independent from each other and equal in number to the new parameters, these new invariants can 
be matched to the 5 extra complex phases of $C_{uRk}$, 
and $C_{dRk}$, making them 
independent from each other and from the previous ones.

As one can see, there are $10 + 36(N - 1)$
independent flavour invariants,
thus matching the number of parameters identified at the end of
Sec.~\ref{sec: Physical Parameters}.
The invariants were introduced in a systematic way. 
First, invariants were constructed to match the quark masses and eigenvalues of the interactions 
with $\mathcal{H}_k$. Then, new independent invariants were constructed that also allowed the 
correspondence with the mixing angles and CKM-like phase of the mixing matrices introduced in 
section \ref{sec: The idea}. After that, new independent invariants were constructed that also 
allowed the correspondence with the other 2 complex phases of both $C_{uLk}$, and $C_{dLk}$. 
Finally, new independent invariants were constructed that also 
allowed the correspondence with the other 5 complex phases of both $C_{uRk}$, and $C_{dRk}$.
Each step requires the previous one in order to guarantee that the new invariants are independent 
from the ones already constructed, and from each other. 

Lastly, one can construct a myriad of other invariants. But since all the physical parameters 
have been matched to the invariants presented here, and since the algebraic expression of said 
new invariants would have to be parametrised by the physical parameters, they would therefore 
also have to be parametrised by the listed invariants, making any new invariant not independent.

The listed invariants are all independent and same 
in number to the physical parameters. 
Therefore, up to sign information, they constitute a basis.

It should also be pointed out that this is only valid in the case
where no parameter in the Lagrangian vanishes and there are no degeneracies.

\section{\label{sec:Conclusions}Conclusions}
In this work, a systematic way of analysing the flavour sector of the NHDM was constructed.
A number of ``mixing matrices'' (akin to the CKM matrix) were introduced,
including the precise way to make them physical by evoking the charged Higgs basis.
The number of magnitudes and phases was computed in a variety of ways.
The example of the 2HDM was also discussed, found to be consistent with
previous ways of analysing the same physics, and expanded to include a complete set of
flavour basis invariants.

The new parameters of the NHDM were also obtained in terms of WBT
invariant traces of flavour matrices.
This result will be useful for physical basis independent studies of the flavour puzzle.
It will also be important for studies of the renormalization group evolution.
As is well known, writing the renormalization group equations in a
specific basis has to contend with the basis rotation occurring during the evolution.
The invariants developed here can be used in basis independent scale evolution,
as outlined for the SM in \cite{Bento:2023owf}.

The physical parameters developed here to fully describe flavour in a NHDM,
and their expression in terms of invariants, will be of use in formulating
flavour studies directly in terms of physical quantities.

\section*{Acknowledgements}
We are grateful to F.~J. Botella for reading and commenting on this manuscript.
This work is supported in part by the Portuguese
Fundação para a Ciência e Tecnologia (FCT) through the PRR (Recovery and Resilience
Plan), within the scope of the investment "RE-C06-i06 - Science Plus Capacity Building", 
measure "RE-C06-i06.m02 - Reinforcement of financing for International Partnerships in Science,
Technology and Innovation of the PRR", under the project with reference 2024.01362.CERN,
and through Contracts UIDB/00777/2020, and UIDP/00777/2020,
partially funded through POCTI (FEDER), COMPETE, QREN, and the EU.

\appendix

\section{\label{sec: unitary matrix}3 by 3 Unitary Matrix Parametrisation}
Let us define
\begin{equation}
\textbf{R}_{12}
=
 \begin{pmatrix}
c_{12} & s_{12} & 0 \\
-s_{12} & c_{12} & 0 \\
0 & 0 & 1
\end{pmatrix}\, ,    
\end{equation}
where, henceforth, $c_{ij}=\cos(\theta_{ij})$,
$s_{ij}=\sin(\theta_{ij})$,
and 
$\textbf{R}_{-12}=\textbf{R}^T_{12}$.
Moreover,
we define
\begin{eqnarray}
\{a,b,c\} &=& \textrm{diag} \{e^{ia},e^{ib},e^{ic}\}\, ,
\nonumber\\
\{1,d,f\} &=& \textrm{diag} \{1,e^{id},e^{if}\}\, ,    
\end{eqnarray}
and similarly for extra ``$1$''s.

A general 3 by 3 unitary matrix can be parametrised 
in terms of 3 mixing angles and 6 complex phases as \cite{Rasin:1997pn}
\begin{eqnarray}
U_{uni}
&=&
\begin{pmatrix}
1 & 0 & 0 \\
0 & e^{i\alpha_1} & 0 \\
0 & 0 & e^{i\alpha_2}
\end{pmatrix} \begin{pmatrix}
1 & 0 & 0 \\
0 & c_{23} & s_{23} \\
0 & -s_{23} & c_{23}
\end{pmatrix}  
\nonumber\\
&\times&
\begin{pmatrix}
e^{-i\delta} & 0 & 0 \\
0 & 1 & 0 \\
0 & 0 & 1
\end{pmatrix} \begin{pmatrix}
c_{13} & 0 & s_{13} \\
0 & 1 & 0 \\
-s_{13} & 0 & c_{13}
\end{pmatrix}  \begin{pmatrix}
e^{i\delta} & 0 & 0 \\
0 & 1 & 0 \\
0 & 0 & 1
\end{pmatrix}
\nonumber\\
&\times&
 \begin{pmatrix}
c_{12} & s_{12} & 0 \\
-s_{12} & c_{12} & 0 \\
0 & 0 & 1
\end{pmatrix} \begin{pmatrix}
e^{i\alpha_3} & 0 & 0 \\
0 & e^{i\alpha_4} & 0 \\
0 & 0 & e^{i\alpha_5}
\end{pmatrix}\, .
\label{unitary parametrisation}
\end{eqnarray}
For a more helpful notation,
we start by denoting the second, fourth, and sixth matrices, respectively by
$\textbf{R}_{23}$, $\textbf{R}_{13}$, and $\textbf{R}_{12}$.
Then,
\begin{eqnarray}
U_{uni} &=& \{1,\alpha_1,\alpha_2\}\textbf{R}_{23}
\{-\delta,1,1\}\textbf{R}_{13}\{\delta,1,1\}
\textbf{R}_{12}\{\alpha_3,\alpha_4,\alpha_5\}\, .
\nonumber\\
&&
\label{U2}
\hspace{1cm}
\end{eqnarray}
%

%Of notice is the fact that the 4 middle matrices make the standard
%parametrisation of the CKM matrix. As will be shown below, this is no coincidence.

\section{\label{app: NHDM}Algebraic Expressions of Flavour Invariants for the NHDM}
Using the cyclic property of the trace, the SM flavour invariants simplify to
\begin{eqnarray}
\label{simplification invariants SM}
&&
I_1^0 = Tr(D_u^2), \; \; \; \; I_2^0 = Tr(D_u^4) \; \; \; \; I_3^0 = Tr(D_u^6),
\nonumber\\
&&
I_4^0 = Tr(D_d^2), \; \; \; \; I_5^0 = Tr(D_d^4) \; \; \; \; I_6^0 = Tr(D_d^6),
\nonumber\\
&&
I_7^0 = Tr(D_u^2VD_d^2V^\dag), \; \; \; \; I_8^0 = Tr(D_u^4VD_d^2V^\dag), 
\nonumber\\
&&
I_9^0 = Tr(D_u^2VD_d^4V^\dag), \; \; \; \; I_{10}^0 = Tr(D_u^4VD_d^4V^\dag).
\end{eqnarray}
The flavour invariants involving the eigenvalues of the interaction with $\mathcal{H}_k$ simplify to
\begin{eqnarray}
\label{simplification invariants eigenvalues}
&&
I_{1}^k = Tr(E_{uk}^2), \; \; \; \;I_{2}^k = Tr(E_{uk}^4) \; \; \; \; I_{3}^k = Tr(E_{uk}^6),
\nonumber\\
&&
I_{4}^k = Tr(E_{dk}^2), \; \; \; \; I_{5}^k = Tr(E_{dk}^4) \; \; \; \; I_{6}^k = Tr(E_{dk}^6).
\end{eqnarray}
The flavour invariants used to obtain the 3 mixing angles and
CKM-like phase of the matrices $C_{uLk}$, $C_{dLk}$, $C_{uRk}$, and $C_{dRk}$
simplify to
\begin{eqnarray}
\label{simplification invariants angles}
&&
I_{7}^k = Tr(D_u^2C_{uLk}E_{uk}^2C_{uLk}^\dag) \; \; \; \; I_{8}^k = Tr(D_u^4C_{uLk}E_{uk}^2C_{uLk}^\dag), 
\nonumber\\
&&
I_{9}^k = Tr(D_u^2C_{uLk}E_{uk}^4C_{uLk}^\dag)
\; \; \; \; I_{10}^k = Tr(D_u^4C_{uLk}E_{uk}^4C_{uLk}^\dag), 
\nonumber\\
&&
I_{11}^k = Tr(D_d^2C_{dLk}E_{dk}^2C_{dLk}^\dag)
\; \; \; \; I_{12}^k = Tr(D_d^4C_{dLk}E_{dk}^2C_{dLk}^\dag), 
\nonumber\\
&&
I_{13}^k = Tr(D_d^2C_{dLk}E_{dk}^4C_{dLk}^\dag)
\; \; \; \; I_{14}^k = Tr(D_d^4C_{dLk}E_{dk}^4C_{dLk}^\dag), 
\nonumber\\
&&
I_{15}^k = Tr(D_u^2C_{uRk}E_{uk}^2C_{uRk}^\dag)
\; \; \; \; I_{16}^k = Tr(D_u^4C_{uRk}E_{uk}^2C_{uRk}^\dag), 
\nonumber\\
&&
I_{17}^k = Tr(D_u^2C_{uRk}E_{uk}^4C_{uRk}^\dag)
\; \; \; \; I_{18}^k = Tr(D_u^4C_{uRk}E_{uk}^4C_{uRk}^\dag), 
\nonumber\\
&&
I_{19}^k = Tr(D_d^2C_{dRk}E_{dk}^2C_{dRk}^\dag)
\; \; \; \; I_{20}^k = Tr(D_d^4C_{dRk}E_{dk}^2C_{dRk}^\dag), 
\nonumber\\
&&
I_{21}^k = Tr(D_d^2C_{dRk}E_{dk}^4C_{dRk}^\dag)
\; \; \; \; I_{22}^k = Tr(D_d^4C_{dRk}E_{dk}^4C_{dRk}^\dag).
\nonumber\\
&&
\end{eqnarray}
The flavour invariants used to obtain the $C_{uLk}$, and $C_{dLk}$ extra phases simplify to
\begin{eqnarray}
\label{simplification invariants left phases}
&&
I_{23}^k = Tr(E_{uk}^2C_{uLk}^\dag VD_d^2V^\dag C_{uLk}),
\nonumber\\
&&
I_{24}^k = Tr(E_{uk}^4C_{uLk}^\dag VD_d^2V^\dag C_{uLk}),
\nonumber\\
&&
I_{25}^k = Tr(E_{dk}^2C_{dLk}^\dag V^\dag D_u^2V C_{dLk}),
\nonumber\\
&&
I_{26}^k = Tr(E_{dk}^4C_{dLk}^\dag V^\dag D_u^2V C_{dLk}).
\end{eqnarray}
The flavour invariants used to obtain the $C_{uRk}$, and $C_{dRk}$ extra phases simplify to
\begin{eqnarray}
\label{simplification invariants right phases}
&&
I_{27}^k = \Re{\left(Tr(D_u C_{uRk} E_{uk} C_{uLk}^\dag V D_d C_{dRk} E_{dk} C_{dLk}^\dag V^\dag)\right)},
\nonumber\\
&&
I_{28}^k = \Im{\left(Tr(D_u C_{uRk} E_{uk} C_{uLk}^\dag V D_d C_{dRk} E_{dk} C_{dLk}^\dag V^\dag)\right)},
\nonumber\\
&&
I_{29}^k = \Re{\left(Tr(D_u^3 C_{uRk} E_{uk} C_{uLk}^\dag V D_d C_{dRk} E_{dk} C_{dLk}^\dag V^\dag)\right)},
\nonumber\\
&&
I_{30}^k = \Im{\left(Tr(D_u^3 C_{uRk} E_{uk} C_{uLk}^\dag V D_d C_{dRk} E_{dk} C_{dLk}^\dag V^\dag)\right)},
\nonumber\\
&&
I_{31}^k = \Re{\left(Tr(D_u C_{uRk} E_{uk} C_{uLk}^\dag V D_d^3 C_{dRk} E_{dk} C_{dLk}^\dag V^\dag)\right)},
\nonumber\\
&&
I_{32}^k = \Im{\left(Tr(D_u C_{uRk} E_{uk} C_{uLk}^\dag V D_d^3 C_{dRk} E_{dk} C_{dLk}^\dag V^\dag)\right)},
\nonumber\\
&&
I_{33}^k = \Re{\left(Tr(D_u C_{uRk} E_{uk}^3 C_{uLk}^\dag V D_d C_{dRk} E_{dk} C_{dLk}^\dag V^\dag)\right)},
\nonumber\\
&&
I_{34}^k = \Im{\left(Tr(D_u C_{uRk} E_{uk}^3 C_{uLk}^\dag V D_d C_{dRk} E_{dk} C_{dLk}^\dag V^\dag)\right)},
\nonumber\\
&&
I_{35}^k = \Re{\left(Tr(D_u C_{uRk} E_{uk} C_{uLk}^\dag V D_d C_{dRk} E_{dk}^3 C_{dLk}^\dag V^\dag)\right)},
\nonumber\\
&&
I_{36}^k = \Im{\left(Tr(D_u C_{uRk} E_{uk} C_{uLk}^\dag V D_d C_{dRk} E_{dk}^3 C_{dLk}^\dag V^\dag)\right)}.
\nonumber\\
&&
\end{eqnarray}
Then, for a diagonal matrix $D = \textrm{diag} (m_1, m_2, m_3)$, 
\begin{equation}
\label{invariants expression eigenvalues}
    Tr(D^p) = m_1^p + m_2^p + m_3^p\, ,
\end{equation}
which allows one to write invariants $I_1^0$ through $I_6^0$ in terms of the quark masses
and invariants $I_1^k$ through $I_6^k$ in terms of the eigenvalues of the
interaction with $\mathcal{H}_k$.

Also, for diagonal matrices $D = \textrm{diag} (m_1, m_2, m_3)$,
$E = \textrm{diag} (n_1, n_2, n_3)$, and a general unitary matrix $U_{uni}$,
parametrised as Eq.~\eqref{unitary parametrisation},
\begin{eqnarray}
\label{invariants expression angles}
&&
Tr(D^pU_{uni}E^qU_{uni}^\dag) =  m_1^pn_1^qc_{12}^2c_{13}^2
+ m_3^pn_3^qc_{13}^2c_{23}^2
\nonumber\\
&&
+ m_1^pn_2^qc_{13}^2s_{12}^2 + m_1^pn_3^qs_{13}^2
+ m_2^pn_3^qc_{13}^2s_{23}^2
\nonumber\\
&&
+ m_3^pn_2^q \left( c_{12}^2s_{23}^2 + 2c_{12}s_{12}s_{13}c_{23}s_{23}c_\delta
+ s_{12}^2s_{13}^2c_{23}^2 \right)
\nonumber\\
&&
+ m_3^pn_1^q \left( s_{12}^2s_{23}^2 - 2c_{12}s_{12}s_{13}c_{23}s_{23}c_\delta
+ c_{12}^2s_{13}^2c_{23}^2 \right)
\nonumber\\
&&
+ m_2^pn_1^q \left( s_{12}^2c_{23}^2 + 2c_{12}s_{12}s_{13}c_{23}s_{23}c_\delta
+ c_{12}^2s_{13}^2s_{23}^2 \right)
\nonumber\\
&&
+ m_2^pn_2^q \left( c_{12}^2c_{23}^2 - 2c_{12}s_{12}s_{13}c_{23}s_{23}c_\delta
+ s_{12}^2s_{13}^2s_{23}^2 \right),
\nonumber\\
&&
\end{eqnarray}
which allows one to write invariants $I_7^0$ through $I_{10}^0$ in terms of the quark masses
and CKM parameters, and invariants $I_7^k$ through $I_{22}^k$ in terms of the quark masses,
eigenvalues of the interaction with $\mathcal{H}_k$, and mixing angles and 
CKM-like phase of the respective $C_k$ matrix.
%%%%%

Then, for diagonal matrices $D = \textrm{diag} (m_1, m_2, m_3)$, $E = \textrm{diag} (n_1, n_2, n_3)$,
a CKM{-like unitary matrix $Y_{uni}$ with only one complex phase,
and unitary matrices $U_{uni}$ and $V_{uni}$ without the three right most phases
of Eq.~\eqref{unitary parametrisation},
\begin{eqnarray}
\label{invariants expression left phases 1}
&&
Tr(E^pU_{uni}^\dag Y_{uni} D^q Y_{uni}^\dag U_{uni}) =  
\nonumber\\
&&
\sum_{i,j=1}^{3} n_i^pm_j^q|(U_{uni}^\dag Y_{uni})_{ij}|^2\, ,
%=
%\nonumber\\
%&&
%n_1^pm_1^q |(U_{uni}^\dag Y_{uni})_{11}|^2 + n_1^pm_2^q |(U_{uni}^\dag Y_{uni})_{12}|^2
%\nonumber\\
%&&
%+ n_1^pm_3^q |(U_{uni}^\dag Y_{uni})_{13}|^2 + n_2^pm_1^q |(U_{uni}^\dag Y_{uni})_{21}|^2
%\nonumber\\
%&&
%+ n_2^pm_2^q |(U_{uni}^\dag Y_{uni})_{22}|^2 + n_2^pm_3^q |(U_{uni}^\dag Y_{uni})_{23}|^2
%\nonumber\\
%&&
%+ n_3^pm_1^q |(U_{uni}^\dag Y_{uni})_{31}|^2 + n_3^pm_2^q |(U_{uni}^\dag Y_{uni})_{32}|^2
%\nonumber\\
%&&
%+ n_3^pm_3^q |(U_{uni}^\dag Y_{uni})_{33}|^2
\end{eqnarray}
\begin{eqnarray}
\label{invariants expression left phases 2}
&&
Tr(E^pV_{uni}^\dag Y_{uni}^\dag D^q Y_{uni} V_{uni}) =  
\nonumber\\
&&
\sum_{i,j=1}^{3} n_i^pm_j^q|(V_{uni}^\dag Y_{uni}^\dag)_{ij}|^2\, ,
%\nonumber\\
%&&
%n_1^pm_1^q |(V_{uni}^\dag Y_{uni}^\dag)_{11}|^2 + n_1^pm_2^q |(V_{uni}^\dag Y_{uni}^\dag)_{12}|^2
%\nonumber\\
%&&
%+ n_1^pm_3^q |(V_{uni}^\dag Y_{uni}^\dag)_{13}|^2 + n_2^pm_1^q |(V_{uni}^\dag Y_{uni}^\dag)_{21}|^2
%\nonumber\\
%&&
%+ n_2^pm_2^q |(V_{uni}^\dag Y_{uni}^\dag)_{22}|^2 + n_2^pm_3^q |(V_{uni}^\dag Y_{uni}^\dag)_{23}|^2
%\nonumber\\
%&&
%+ n_3^pm_1^q |(V_{uni}^\dag Y_{uni}^\dag)_{31}|^2 + n_3^pm_2^q |(V_{uni}^\dag Y_{uni}^\dag)_{32}|^2
%\nonumber\\
%&&
%+ n_3^pm_3^q |(V_{uni}^\dag Y_{uni}^\dag)_{33}|^2
\end{eqnarray}
which, if expanded, would allow one to write invariants $I_{23}^k$ through $I_{26}^k$ in terms
of the quark masses, eigenvalues of the interaction with $\mathcal{H}_k$, mixing angles and 
phase of the CKM matrix, and mixing angles and phases of the respective $C_{Lk}$ matrix.

Finally, for diagonal matrices $D = \textrm{diag} (m_1, m_2, m_3)$,
$E = \textrm{diag} (n_1, n_2, n_3)$, $F = \textrm{diag} (a_1, a_2, a_3)$,
$G = \textrm{diag} (b_1, b_2, b_3)$, a CKM-like unitary matrix $Y_{uni}$
with only one complex phase, unitary matrices $U_{uni}$ and $V_{uni}$
without the three right most phases,
and arbitrary unitary matrices $W_{uni}$ and $X_{uni}$,
\begin{eqnarray}
\label{invariants expression right phases}
&&
Tr(D^pW_{uni}E^qU_{uni}^\dag Y_{uni}F^rX_{uni}G^tV_{uni}^\dag Y_{uni}^\dag) = 
\nonumber\\
&&
\sum_{i,j,k,l=1}^{3}
m_i^pn_j^qa_k^rb_l^t(W_{uni})_{ij}(U_{uni}^\dag Y_{uni})_{jk}(X_{uni})_{kl}(V_{uni}^\dag Y_{uni}^\dag)_{li}.
\nonumber\\
&&
\label{huge}
\end{eqnarray}
Taking the real and imaginary parts and expanding would allow one to write 
invariants $I_{27}^k$ through $I_{36}^k$ in terms of the quark masses, eigenvalues of the 
interaction with $\mathcal{H}_k$, and mixing angles and phases of the CKM matrix and the 
respective $C_{uLk}$, $C_{dLk}$, $C_{uRk}$, $C_{dRk}$ matrices.

\bibliographystyle{elsarticle-num} 
\bibliography{refs}



%% For example:

%% \bibitem[Aladro et al.(2015)]{Aladro15} Aladro, R., Martín, S., Riquelme, D., et al. 2015, \aas, 579, A101

%%\end{thebibliography}

\end{document}